\documentclass{aa}
\usepackage[varg]{txfonts}

\usepackage{array, multirow, graphicx,natbib,txfonts,color,tabularx,amsmath}
\usepackage{lscape,float,placeins}
\usepackage{hyperref}
\usepackage{afterpage}
\usepackage{threeparttable}
\bibpunct{(}{)}{;}{a}{}{,}

\defcitealias{Campbell2013}{C13}
\defcitealias{Lapenna2016}{L16}

\begin{document}

\title{First evidence of multiple populations along the AGB from Str\"omgren photometry \thanks{Based on observations made with the Isaac Newton Telescope operated on the island of La Palma by the Isaac Newton Group in the Spanish Observatorio del Roque de los Muchachos of the Instituto de Astrof\'{\i}sica de Canarias.}\,\thanks{Tables B.1 and B.2 are only available in electronic form at the CDS via anonymous ftp to cdsarc.u-strasbg.fr (130.79.128.5) or via \url{http://cdsweb.u-strasbg.fr/cgi-bin/qcat?J/A+A/}}  }

\author{Pieter Gruyters\inst{1} \and Luca Casagrande\inst{2} \and Antonino P. Milone\inst{2} \and Simon T. Hodgkin\inst{3} \and Aldo Serenelli\inst{4} \and Sofia Feltzing\inst{1}}

\offprints{pieter.gruyters@physics.uu.se}

\institute{Lund Observatory, Department of Astronomy and Theoretical Physics, Box 43, SE-221 00 Lund, Sweden \and Research School of Astronomy and Astrophysics, Mount Stromlo Observatory, The Australian National University, ACT 2611, Australia \and Institute of Astronomy, Madingley Road, Cambridge CB3 0HA, UK \and Institute of Space Sciences (IEEC-CSIC), Carrer de Can Magrans S/N, 08193 Barcelona, Spain}

\date{Received  / Accepted}

\authorrunning{P. Gruyters}
\titlerunning{Multiple pop on AGB}

%*****************************************************************************
%                      ABSTRACT
%*****************************************************************************
\abstract
{Spectroscopic studies have demonstrated that nearly all Galactic globular clusters (GCs) harbour multiple stellar populations with different chemical compositions.  Moreover, colour-magnitude diagrams based exclusively on Str\"omgrem photometry have allowed us to identify and characterise multiple populations along the RGB of a large number of clusters.
In this paper we show for the first time that {\bf Str\"omgren photometry is also very efficient at identifying multiple populations  along the AGB, and demonstrate that the AGB} of M\,3, M\,92, NGC\,362, NGC\,1851, and NGC\,6752 are not consistent with a single stellar population. We also provide a catalogue of RGB and AGB stars photometrically identified in these clusters for further spectroscopic follow-up studies. We combined photometry and elemental abundances from the literature for RGB and AGB stars in NGC\,6752 where the presence of multiple populations along the AGB has been widely debated. We find that, while the MS, SGB, and RGB host three stellar populations with different helium and light element abundances, only two populations of AGB stars are present in the cluster. 
  {\bf These results are consistent with standard evolutionary theory.}
 }

% Max 6 keywords!

\maketitle

%******************************************************************
%                    INTRODUCTION
%******************************************************************

\section{Introduction}\label{sect:intro}
        {\bf Nearly all the old} globular clusters (GCs) are characterised by star-to-star variations in the light elements, {\bf including He, C, N, O, and Na  \citep[e.g.][]{Milone2014b,Pancino2010,Carretta2009a}. Moreover, some clusters exhibit variations in Li, Mg, Al, and Si \citep[e.g.][]{Dorazi2015,Carretta2009b}.} Some massive `anomalous' clusters also have internal variations in some of the heavy elements \citep[e.g.][]{Marino2015};  {\bf see \citet{Gratton2012} for a review on the chemical composition of multiple populations in GCs.}\\

 The origin of multiple stellar populations in GCs is still not fully understood. Some authors have concluded that GCs have experienced multiple episodes of star formation during which a second generation (SG) of stars is formed from material polluted by processed material from massive first generation (FG) stars \citep[e.g.][]{Dantona2002,Decressin2007a,Denissenkov2014}. \\
 
Other groups suggest that the chemical variations are due to accretion of polluted material in the pre-main sequence phase \citep{Bastian2013}; \citet{Renzini2015,Bastian2015}, and \citet{Dantona2016} provide critical discussions of the different scenarios.
 
 Photometry is a powerful tool to identify multiple stellar populations. Multi-wavelength photometry from the {\it Hubble Space Telescope} ({\it HST}) has revealed multiple main sequences (MSs), sub-giant branches (SGBs) and red giant branches (RGBs) in nearly all the analysed GCs \citep[see e.g.][and references therein]{Piotto2015,Milone2017}.   These multiple sequences correspond to stellar populations with different helium and light elements abundances and strongly affect the morphology of the colour-magnitude diagram (CMD) of star clusters. Recent papers have provided direct evidence that even the shape of the horizontal branch is closely connected with the presence of multiple populations \citep[e.g.][and references therein]{Marino2011,Marino2014,Gratton2012} .

Multiple RGBs have also been identified using ground-based photometry \citep[e.g.][]{Marino2008,Monelli2013} as demonstrated by the pioneering papers by Frank Grundahl and his collaborators based on Str\"omgren photometry \citep[e.g.][]{Grundahl1998,Grundahl1999,Yong2008,Carretta2011}.

The asymptotic giant branch (AGB) has been poorly investigated in the context of multiple populations. {\bf Nevertheless, the presence (or the absence) of stellar populations along the AGB is crucial to constrain the models of stellar evolution. Indeed helium-rich stars in GCs, which have lower mass than stars with primordial helium abundance during the horizontal branch (HB) phase, can evolve into so-called AGB manqu{\'e} stars that never reach the AGB \citep[e.g.][]{Sweigart1976,Dorman1993}}.  

{\bf Early spectroscopic studies by \citet{Mallia1978}, \citet{Norris1981}, \citet{Smith1993}, and more recently \citet{Campbell2006} report a possible lack of CN-strong stars along the AGB and therefore hint that second-population stars in GCs avoid the AGB phase. In contrast, recent studies by, for example\citet{Ivans2001,Johnson2015}, and \citet{Wang2016} have revealed that the AGB of some clusters, such as NGC\,5904 and 47\,Tuc, exhibit star-to-star light element variations, in close analogy with what has been observed along their RGBs.}
 
 NGC\,6752 is an intriguing case and has been strongly debated in recent papers. Photometry and spectroscopy of RGB and MS stars have revealed that this cluster hosts three stellar populations with different content of helium and light elements and that only $\sim 30\%$ of its stars belong to a population of FG stars with primordial helium abundance \citep{Yong2013,Milone2013,Milone2017,Dotter2015}; {\bf see also \citet{Carretta2007} for the study of the sodium-oxygen anti-correlation along the RGB of NGC\,6752}.  Intriguingly, \citet{Campbell2013} derived Na abundances for 24 RGB and 20 AGB stars and from the [Na/Fe] distribution they concluded that there are no SG AGB stars with enhanced Na and He \citep[i.e. 2nd and 3rd population in NGC\,6752][]{Milone2013}  and that only $\sim 30\%$ of NGC\,6752 stars undergo the AGB phase. This fact would suggest that the majority of NGC\,6752 stars enhanced in He (i.e. SG stars) do not ascend the AGB \citep[see discussion by][]{Cassisi2014}.
 {\bf Noticeably, NGC\,6752 hosts a very extended HB which is well populated on the blue side of the \citet{Grundahl1999} jump \citep[e.g.\,][]{Brown2016}. Campbell and collaborators suggested that the metal enhancement due to radiative levitation of HB stars hotter than $\sim 11,500$ K may cause enhanced mass loss and push HB stars into the AGB-manqu{\'e} phase.}
 This conclusion was challenged by \citet{Lapenna2016} who re-observed the 20 AGB stars and derived Na and O abundances amongst other. Based on their derived abundances \citet{Lapenna2016} conclude that SG AGB stars with enhanced Na are also present \citep[see also][]{Cassisi2014}.

{\bf  Interestingly, the spectroscopic investigation of 6 AGB stars and 13 RGB stars in M\,62 suggests that its second-generation stars avoid the AGB phase \citep{Lapenna2016}. This cluster exhibits a very extended HB and about 80\% of its stars are highly helium enhanced by $\Delta Y \sim$0.08 with respect to the primordial value \citep{Milone2015c}. These observations are consistent with the scenario where AGB-manqu{\'e} stars have high helium abundance \citep[e.g.\,][]{Gratton2010}.}

{\bf The GC M\,4 is another controversial case. This cluster hosts two main stellar populations with different oxygen and sodium abundance, which correspond to distinct photometric sequences along the RGB and the MS \citep[e.g.\,][]{Marino2008, Milone2014b}. The sodium-rich stellar population is only slightly enhanced in helium with respect to the sodium poor population by $\Delta Y \sim 0.02$ \citep[e.g.\,][]{Villanova2012, Nardiello2015}.

In addition, the HB of M\,4 is well populated on both sides of the RR\,Lyrae instability strip but its stars never exceed $\sim$9,000 K and are not affected by radiative levitation. Moreover, \citet{Marino2011} and \citet{Villanova2012} found that {\bf the blue HB stars} are sodium rich and oxygen poor, while {\bf the red HB} stars have the same chemical composition as the first populations.

     In this context, the conclusion of a recent spectroscopic study of AGB stars that second-population stars of M\,4 may avoid the AGB phase \citep{MacLean2016} challenges the prediction of standard stellar evolution theory.
     Indeed we would expect that the blue HB stars of M\,4, which are only marginally helium enhanced and belong to the blue HB, would evolve into the AGB. Noticeably, the result by MacLean and collaborators is not confirmed by the successive work by \citet{Lardo2017} who have analysed $U, B, V, I$ photometry and have found that the AGB of M\,4 is not consistent with a simple population.
      }
   
   In this paper we study for the first time the AGB of five GCs, namely M\,3, M\,92, NGC\,362, NGC\,1851, and NGC\,6752 using Str\"omgren photometry. We are in particular interested in revealing the presence of multiple populations along the AGB in these clusters. It has been shown that CMDs built with the appropriate combination of ultraviolet and optical HST filters are able to identify multiple populations along the AGB  \citep[e.g.][for NGC\,2808, NGC\,7089, and NGC\,6352]{Milone2015a,Milone2015b,Nardiello2015}.
   {\bf Multiple sequences of AGB stars have also been detected with a combination of wide-band ground-based photometry \citep[e.g\,][]{Monelli2013, GarciaHernandez2015, Lardo2017}.}
   Here we leverage on the combination of Str\"omgren filters to achieve the same goal. In fact, it has already been shown that Str\"omgren photometry is very efficient in identifying stellar populations with different chemical compositions \citep[e.g.][]{Grundahl1998,Yong2008}. 

We also combine information from the spectroscopic analysis by \citet{Campbell2013} and \citet{Lapenna2016} and multi-wavelength Str\"omgrem photometry to investigate multiple populations along the AGB of NGC\,6752 in detail.

%#######################################################################################

%******************************************************************
%                    OBSERVATIONS
%******************************************************************

\section{Observations}\label{sect:obs}
The analysis presented in this paper is based on Str\"omgren photometry. We combine our own data obtained during the St\"omgren survey for Asteroseismology and Galactic Archeology \citep[SAGA;][]{Casagrande2014a} and literature data collected by Grundahl and co-workers. The Str\"omgren $uvby$ system \citep{Stromgren1963} was developed to obtain basic stellar parameters such as effective temperature, surface gravity, and metallicity \citep[see e.g.][and references therein]{Arnadottir2010}. 

\subsection{SAGA data}
The SAGA survey \citep{Casagrande2014a} is based on observations collected with the Wide Field Camera (WFC) at the Isaac Newton Telescope (INT). Although the main goal of SAGA is to obtain Str\"omgren photometry of stars in the Kepler and K2 fields, a number of open and globular clusters are also targeted. Here we used photometry obtained in June 2012 for the globular cluster M3. For the data reduction we used an update of the procedure described in \citet{Casagrande2014a} {\bf and derived PSF photometry with the method and computer programs by \citet{Anderson2006a} and \citet{Anderson2006b} adapted to the INT/WFC camera.}

\subsection{Literature data}
We used the photometry collected by Grundahl and co-workers and presented by \citet{Calamida2007}\footnote{The catalogues were downloaded from the web page \url{http://www.oa-roma.inaf.it/spress/gclusters.html}}. The data consists of fully calibrated PSF and aperture photometry. Information concerning the observations, data reduction and calibration procedures can be found in \citet{Grundahl1999} and \citet{Grundahl2002}. We use the data for four GCs, i.e. M\,92, NGC\,362, NGC\,1851, and NGC\,6752. 
%

%#######################################################################################

%******************************************************************
%                    ANALYSIS
%******************************************************************
\section{Analysis}
In this section we analyse the available dataset to investigate multiple stellar populations of five clusters along the AGB and the RGB. In the following subsection we use the example of NGC\,6752 to demonstrate that Str\"omgren photometry can be used to identify multiple populations along the AGB and RGB, while in the Sect.~3.2 we extend the analysis to M3, M92, NGC\,362, and NGC\,1851.

\subsection{Multiple stellar populations in NGC\,6752}
           
The identification and characterisation of multiple populations in the CMD requires accurate photometry.
To obtain the best possible cluster photometry, we have identified a sample of well-measured GC stars, according to three selection criteria, and we limited our analysis to the selected stars.

\begin{itemize}
\item    We first rejected stars in the central regions of each cluster where the photometry has a poorer quality due to stellar crowding. Specifically, we rejected stars with radial distance from the cluster centre, R$<$45 arcsec in the case of M\,3 and stars with R$<$90 arcsec in the remaining clusters;
  
\item We selected stars with relatively small photometric errors, which have been estimated as the ratio between the r.m.s. of the $N$ independent magnitude measurements divided by the square root of $N-1$ ($N\sim10^4-10^5$). Specifically, we rejected all the stars with an error larger than $err_{\rm med}+0.015$, where $err_{\rm med}$ is the median error of stars with similar magnitude (typically $u_{\rm err}<0.01$ for $y<16$).

\item Moreover, we included in our analysis only stars within a narrow $m_{1}=(v-b)-(b-y)$ interval. As the Str\"omgren $m_1$ index is primarily sensitive to metallicity, this criterium ensures that we mostly selected evolved stars with the same metallicity as the cluster and we minimised contamination by field stars.  
\end{itemize}

\begin{figure}
\begin{center}
  \includegraphics[width=0.99\columnwidth]{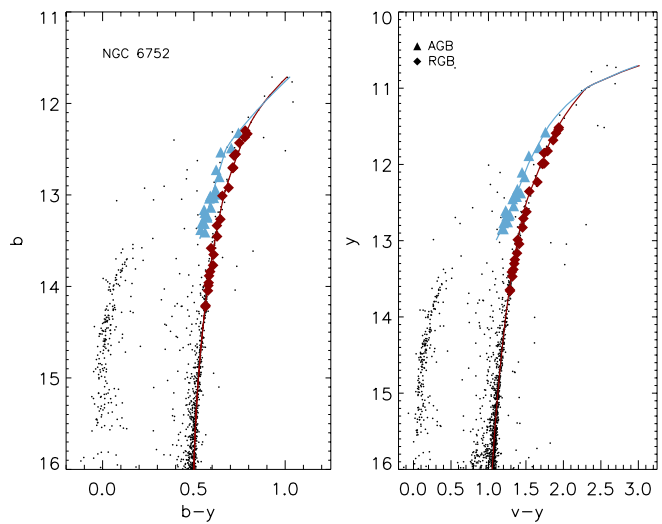}
\caption{{\bf Colour-Magnitude diagrams of ($b$, $b-y$) (left) and ($y$, $v-y$) (right) for NGC\,6752. The RGB and AGB stars studied by \citet{Campbell2013} are denoted with diamonds and triangles, respectively. The blue and brown lines are the fiducial sequences of the AGB and the RGB, respectively.}}\label{Fig:NGC6752_CMD}
\end{center}
\end{figure}

The ($b$, $b-y$) {\bf and ($y$, $v-y$) CMDs} of the selected NGC\,6752 stars are shown in Fig.~\ref{Fig:NGC6752_CMD}. We constructed fiducial sequences for the RGB and AGB by eye {\bf in such a way that they pass through the AGB and RGB stars selected by \citet{Campbell2013}. For the other clusters we used fiducial sequences} to identify a sample of AGB and RGB stars that are located within a certain distance from these fiducial lines. Details on the selection rules are provided in Table\,\ref{Tab:selection} in the appendix.

\begin{figure}
\begin{center}
  \includegraphics[width=0.99\columnwidth]{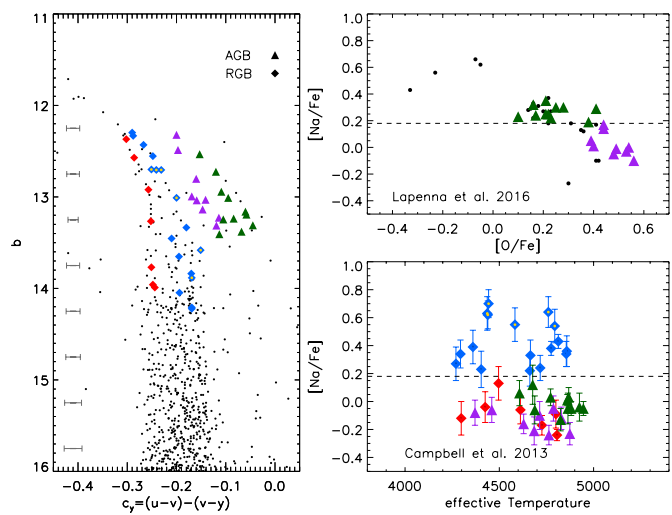}
\caption{{\bf Diagram of ($b$, $c_{\rm y}$) for AGB and RGB stars of NGC\,6752 (left). The typical photometric errors are indicated by the error bars on the left of the diagram. The upper right panel compares the sodium-oxygen anti-correlation for RGB stars (bullets) and AGB stars (triangles) from \citet{Yong2013} and \citet{Lapenna2016}. The dashed line separates population A and B AGB stars, which are coloured green and purple, respectively. The same colours are used to represent these AGB stars in the other panels of this figure. The lower right panel shows [Na/Fe] as a function of the stellar effective temperature for RGB (diamonds) and AGB stars (triangles) from \citet{Campbell2013}. The dashed line separates the two groups first and second population stars defined by Campbell and collaborators, which are coloured red and blue, respectively. RGB stars with [Na/Fe]$>$0.5 are denoted with yellow dots. These symbols and colours are used to represent the same RGB stars in the diagram in the left panel.}}\label{Fig:NGC6752-NGC1851}
\end{center}
\end{figure}

The $c_{\rm y}=(u-v)-(v-y)$ index was defined by \citet{Yong2008} as a colour index that is sensitive to the stellar nitrogen abundance but almost insensitive to temperature. Owing to the N-sensitivity, the $c_{\rm y}$ index can effectively distinguish between stellar populations with different light element content. 
The {\bf left} panel of Fig.~\ref{Fig:NGC6752-NGC1851} shows the ($b$, $c_{\rm y}$) diagram for the selected AGB and RGB stars. The typical observational errors are indicated by the error bars plotted on the left of this diagram.
We immediately note that the $c_{\rm y}$ spread of both AGB and RGB stars is much larger than what we expect from observational errors alone.
To quantify this phenomenon, we determined the spread in the $c_{\rm y}$ index on the AGB and RGB by constructing an RGB and AGB fiducial sequence in the ($y$, $c_{\rm y}$) CMD and calculating the standard deviation around these sequences. We found that the spread in the $c_{\rm y}$ index on the AGB and RGB in NGC\,6752 is $\sim$0.05 mag for AGB and RGB, respectively, and is significantly larger than the spread stemming from the observational error, which is $\sim$0.01 mag.
 
The fact that the RGB is intrinsically broadened in the $c_{\rm y}$ index reveals that NGC\,6752 hosts multiple populations  \citep[see e.g.][]{Yong2008}.
 {\bf Indeed the $c_{\rm y}$ index is strongly sensitive to the nitrogen abundance of the different stellar populations and the $c_{\rm y}$ value of an RGB star strongly correlates with its nitrogen abundance \citep[see Figure 7 from][for an example]{Yong2008}
 Similarly, the $c_{\rm y}$ broadening of AGB stars demonstrates that the AGB is not consistent with a simple population.}

To further investigate stellar populations along the AGB and RGB of NGC\,6752 we combine photometry with elemental abundances from the literature.
In the upper right panel of Fig.~\ref{Fig:NGC6752-NGC1851} we reproduce the sodium-oxygen anti-correlation for AGB stars \citep[coloured triangles;][]{Lapenna2016} and RGB stars \citep[black bullets;][]{Yong2013}.

The RGB of NGC\,6752 hosts three main populations with different O, Na, and helium: the two stellar groups with a helium fraction $Y\sim 0.25-0.26$ are clustered around [O/Fe]$\sim$0.4 and [O/Fe]$\sim$0.2 in the [Na/Fe] versus \,[O/Fe] plane. The third group of stars has extreme oxygen and sodium abundances and has high helium abundances {\bf $Y \sim 0.27-0.28$ \citep{Milone2013,Dotter2015}}. {\bf \citet{Milone2013} called the stellar group with high helium content population C, while the other two populations with $Y \sim 0.25$ and $\sim 0.26$ are indicated as A and B, respectively. For clearness, in the following we adopt the terminology by Milone and collaborators.  Since stars of the same age but different helium abundances have different masses, their presence or absence along the AGB are crucial to constrain stellar evolution models.}

{\bf The RGB and AGB of NGC\,6752 have been studied spectroscopically by \citet{Campbell2013} who discovered that AGB stars span a smaller range of Na than RGB stars. These authors concluded that all second-population stars avoid the AGB phase and invoked extreme mass loss in second-population HB stars. Specifically, they suggested that these stars are located on the blue side of the \citet{Grundahl1999} jump and that their mass loss is enhanced by the effect of metal enhancement associated with radiative levitation.
Campbell and collaborators did not account for the three stellar populations A, B, and C of NGC\,6752, but adopted a simple scheme whereby the analysed stars only include a first population with primordial chemical composition and a second population of stars enhanced in helium and sodium.

{\bf  \citet{Lapenna2016} analysed Na and O of the same AGB stars in NGC\,6752 that were already analysed by \citet{Campbell2013}} and discovered that the AGB hosts stars with different sodium and oxygen abundances, thus demonstrating that the AGB is not consistent with a simple population. The two populations of AGB stars with primordial and intermediate chemical composition, correspond to the populations A and B and have low helium abundance ($Y \lesssim 0.26$). 
 
 Noticeably, Lapenna and collaborators confirmed that the stars with high sodium (hence low oxygen) abundance, are not present along the AGB. These stars belong  to the helium-rich population C identified photometrically}. 
 
{\bf Both populations A and B of AGB stars } are clearly visible in the upper right panel of Fig.~\ref{Fig:NGC6752-NGC1851} and are coloured purple and green, respectively. The same colours are used to represent the same stars in the ($b$, $c_{\rm y}$) diagram of Fig.~\ref{Fig:NGC6752-NGC1851}.

{\bf Population A and population B stars define almost distinct sequences and population B} AGB stars have
 larger $c_{\rm y}$ values than {\bf population A} AGB stars with the same luminosity.
This fact confirms that the $c_{\rm y}$ broadening of AGB stars is actually due to the presence of multiple stellar populations of AGB stars and demonstrates that Str\"omgren photometry is very efficient in identifying multiple populations along the AGB.

In the lower right panel of Fig.~\ref{Fig:NGC6752-NGC1851} we show [Na/Fe] versus $T_{\rm eff}$ for AGB stars (triangles) and RGB stars (diamonds) from \citet{Campbell2013}.  Purple and green colours indicate the AGB stars with primordial and intermediate chemical composition identified in the upper right panel of Fig.~\ref{Fig:NGC6752-NGC1851}. These two groups of stars have, on average, different sodium abundances thus demonstrating that the group of stars that, for simplicity, has been indicated as a first population by \citet{Campbell2013} actually include two groups of stars (A and B) with slightly different helium abundance.

For completeness, we identify in the [Na/Fe] versus $T_{\rm eff}$ plane three groups of RGB stars with different sodium abundances and indicate these with red, blue, and yellow bullets within blue diamonds in Fig.~\ref{Fig:NGC6752-NGC1851}.
We confirm that stars with different sodium abundances populate different RGB regions in the ($b$, $c_{\rm y}$) plane as previously noticed by \citet{Yong2008}. Noticeably, stars with extreme Na and O abundances are not present near the RGB tip. The small number of stars prevents us from making a strong conclusion and further spectroscopic investigation of more stars is needed to determine whether this phenomenon is intrinsic or if it is due to small statistics.

\subsection{Multiple stellar populations along the AGB of M\,3, M\,92, NGC\,362, and NGC\,1851}

Having demonstrated that the $c_{\rm y}$ provides an efficient way to identify multiple populations along the AGB, we extend the analysis to four other clusters {\bf, namely M\,3, M\,92, NGC\,362, and NGC\,1851}.

\begin{figure*}
\begin{center}
\includegraphics[width=0.95\columnwidth]{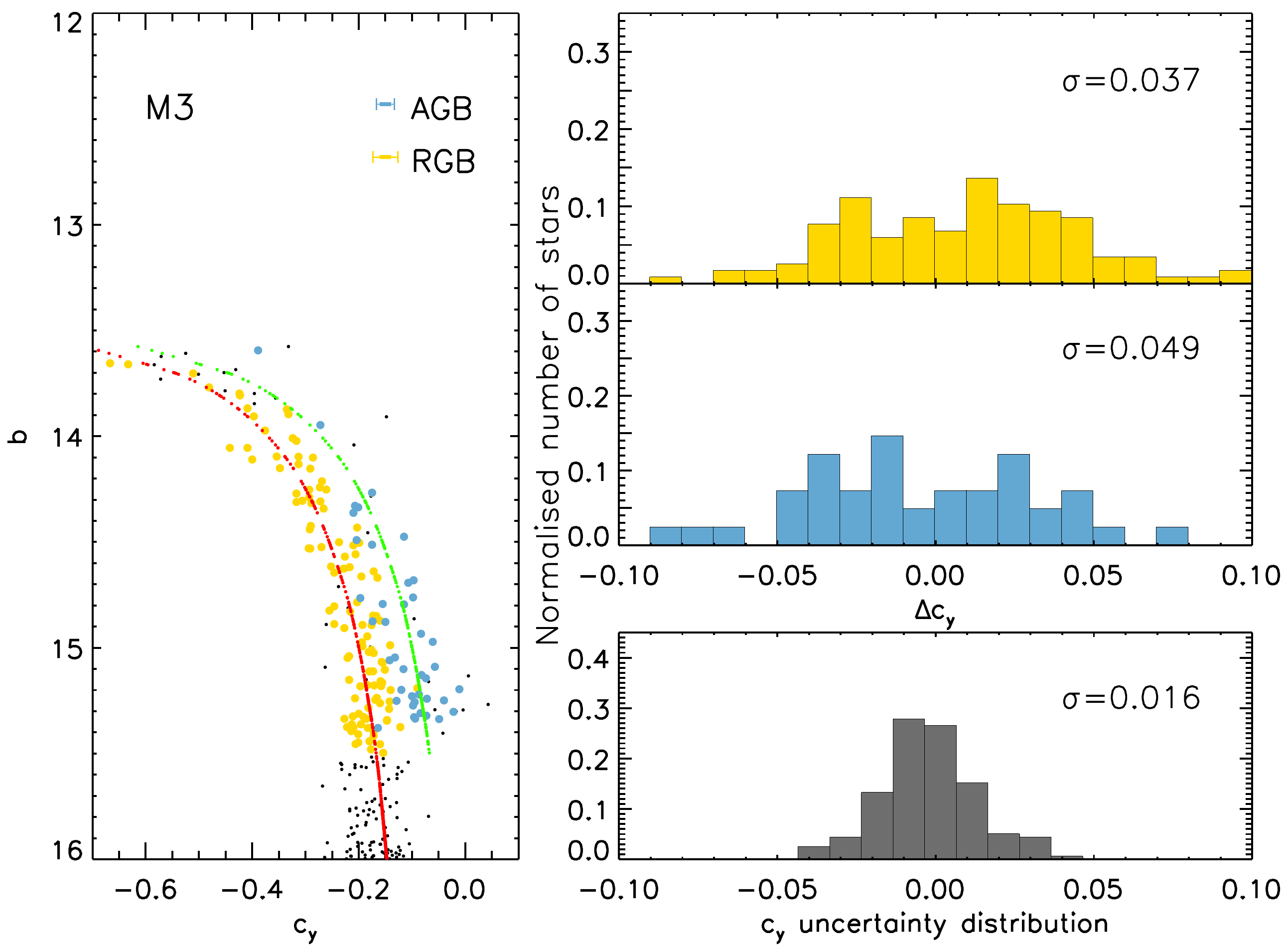} 
\includegraphics[width=0.95\columnwidth]{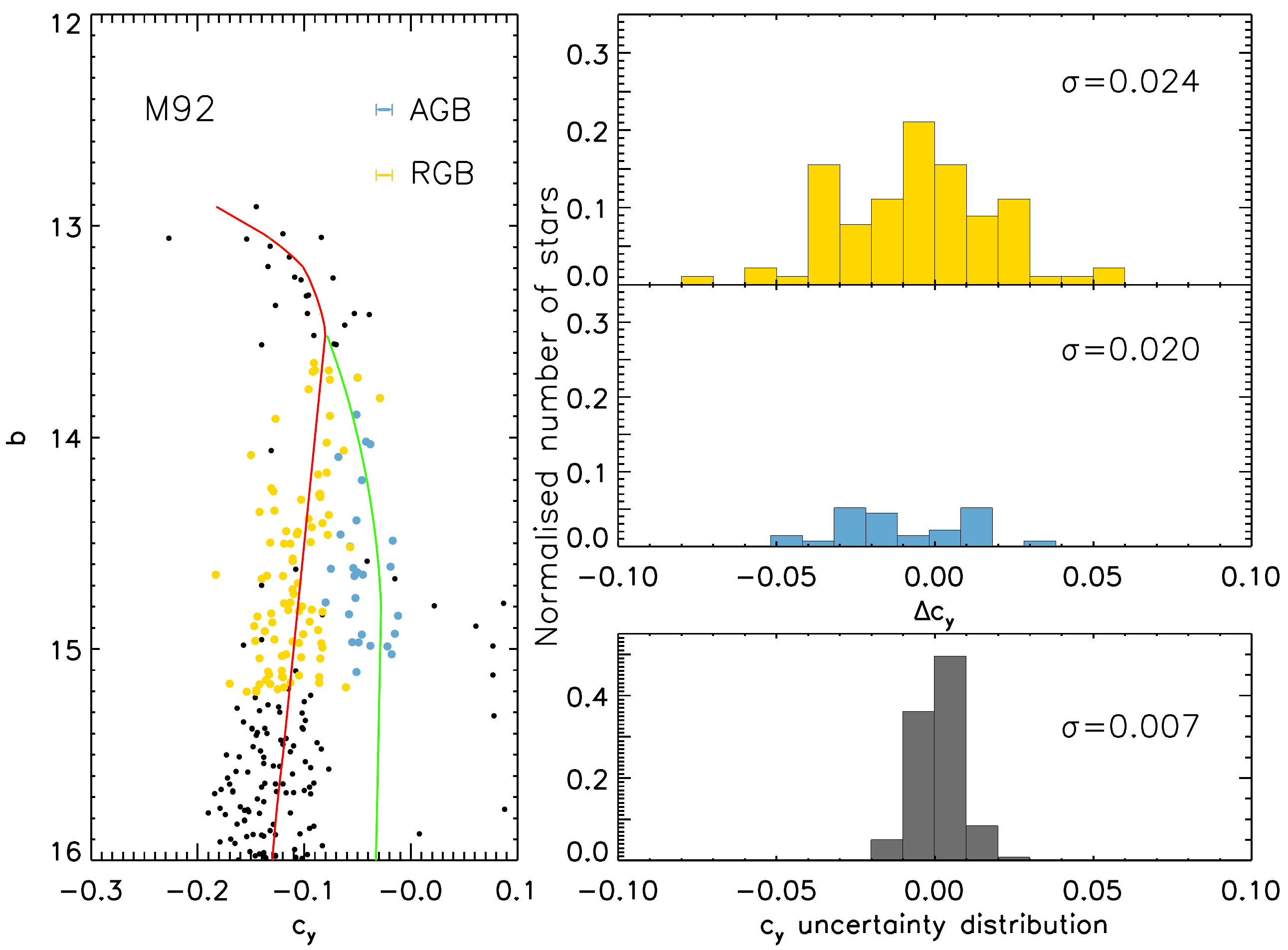} \\
\includegraphics[width=0.95\columnwidth]{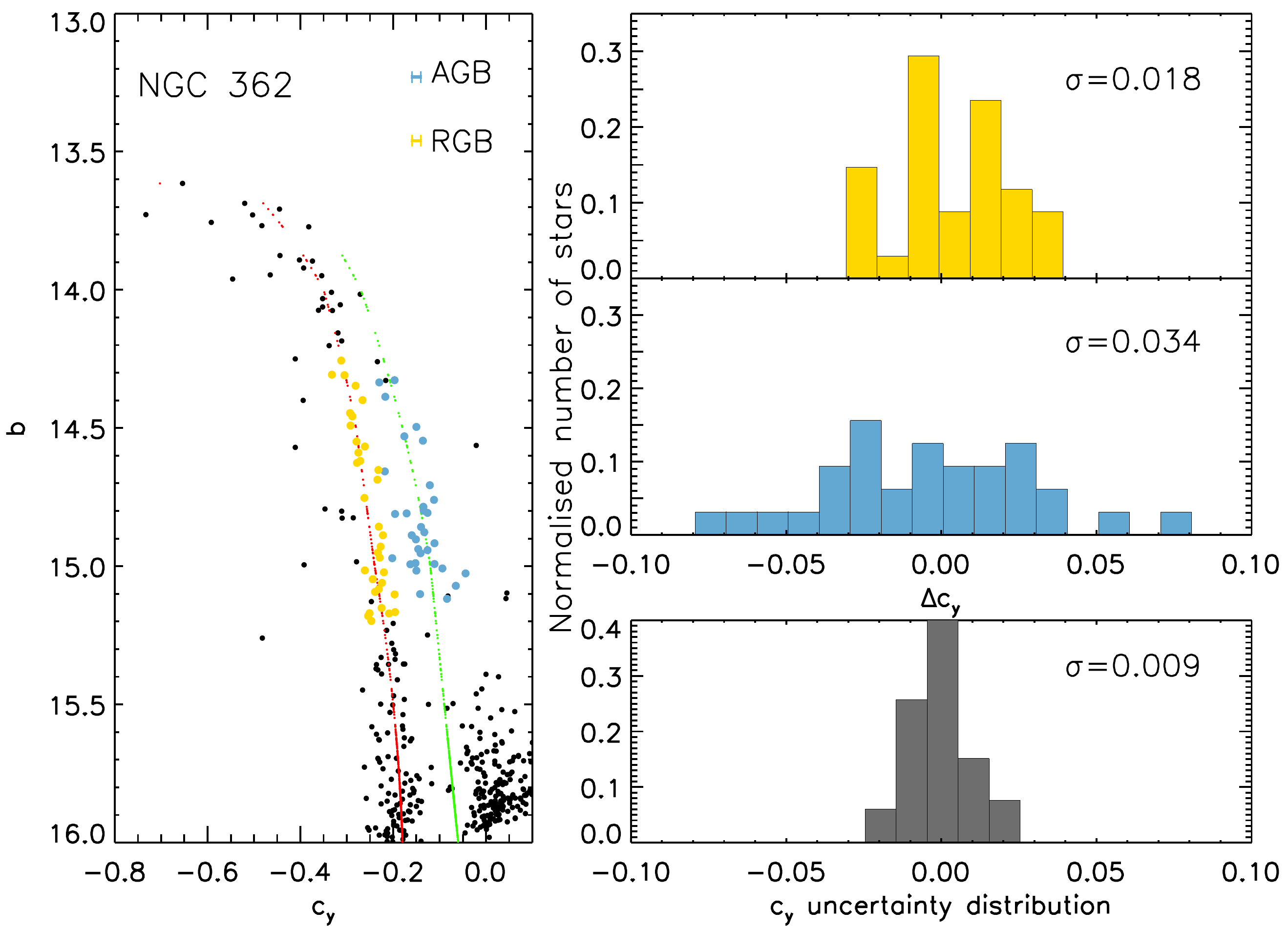}
\includegraphics[width=0.95\columnwidth]{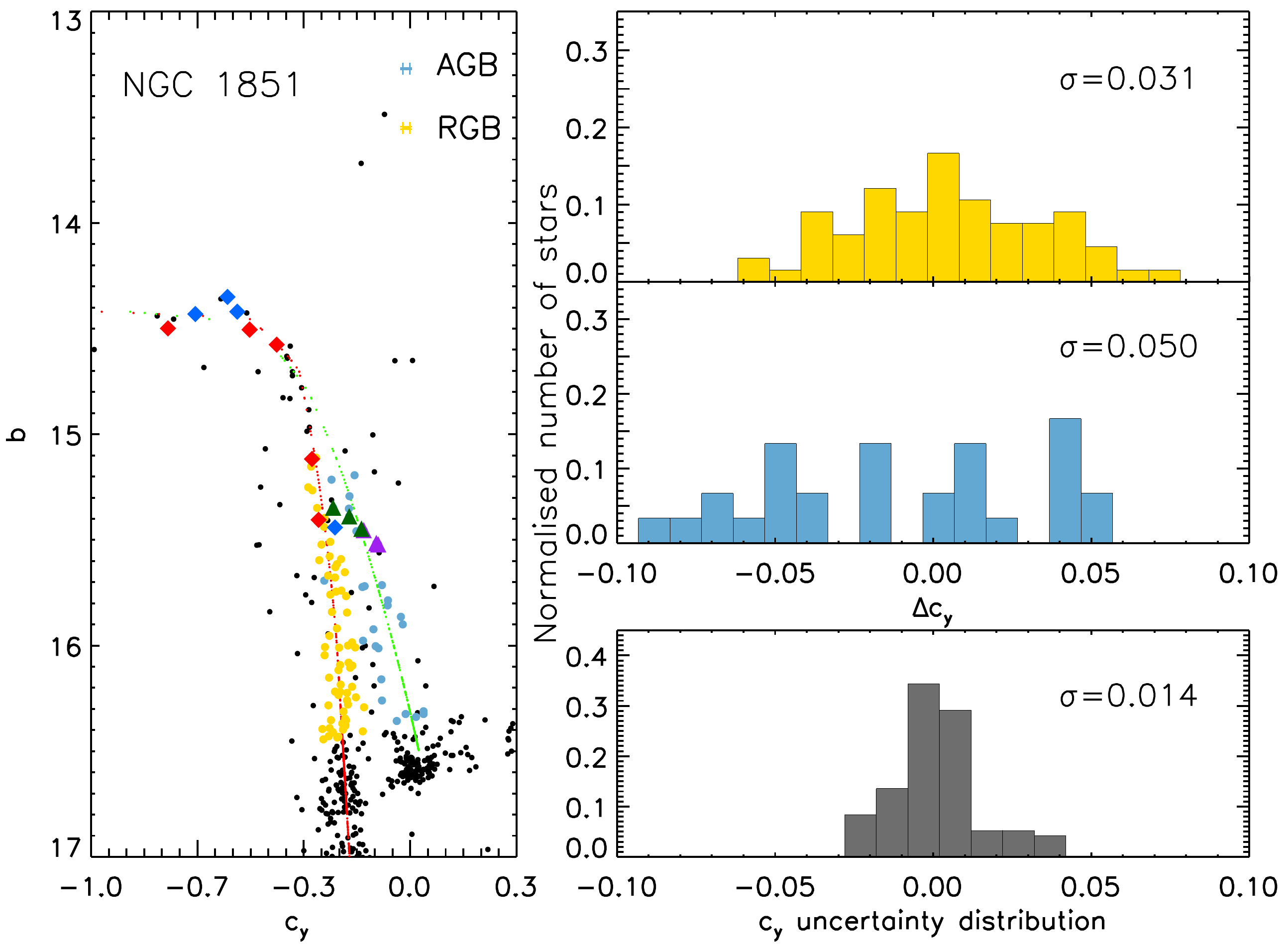} \\
\caption{Colour-magnitude diagrams of $b$ vs.\,$c_{\rm y}$ and $c_{\rm y}$ spreads for M3 (top left), M92 (top right), NGC\,362 (bottom left), and NGC\,1851 (bottom right). In all plots the RGB stars are given by the yellow bullets, while the blue bullets represent the AGB stars. The typical photometric uncertainty for each group is given by the error bars in the upper right part of the CMDs (yellow for RGB and blue for AGB) and the bottom histogram in each plot shows the distribution of the photometric uncertainties in $c_{\rm y}$. The fiducial sequences for the RGB and AGB are also plotted in the CMDs. The coloured histograms show the spread around these fiducial sequences. The standard deviation $\sigma$ is also given. Abundance information for some of the stars in NGC\,1851 is overplotted in the bottom right CMD, where navy blue and purple indicate FG stars and red and green indicate SG stars based on the CNO and Na abundances available in the literature (see text).}\label{Fig:GCs}
\end{center}
\end{figure*}

The results are shown in Fig.\,\ref{Fig:GCs}. The figure shows the $(c_{\rm y},b)$ CMDs of the clusters along with histograms of the spread in $c_{\rm y}$ on the RGB (top panel) and AGB (middle panel) as well as the uncertainty distribution based on the photometric uncertainties in $c_{\rm y}$ (bottom panel). To obtain the spread in $c_{\rm y}$ we construct an RGB and an AGB fiducial sequence (given by the lines in the CMDs) in the ($c_{\rm y},b$) CMD. The actual $c_{\rm y}$ spread is then the standard deviation $\sigma$ around the fiducial sequence. The typical photometric uncertainty on the $c_{\rm y}$ index for the RGB and AGB is also given by the error bar on the right side of the CMDs. The spreads are summarised in Table\,\ref{Tab:cy-spread} along with the average photometric uncertainties in $c_{\rm y}$. \\
Comparing the uncertainty distributions with the spread on the RGB/AGB shows immediately that the spread in $c_{\rm y}$ is real and not the result of the photometric uncertainties. As all clusters have spectroscopically confirmed multiple populations (albeit not on the AGB), we conclude that the spread on the RGB and the AGB is due to the presence of multiple populations. 

{\bf The fact that the AGB of M\,3 is not consistent with a simple stellar population confirms previous findings by \citet{GarciaHernandez2015}. These authors used APOGEE spectroscopy to show that M\,3 second-generation AGB stars with enhanced aluminium abundance. Moreover, the wide spread in that they observed in the $V$ versus $C_{\rm U,B,I}$ further demonstrates that M\,3 hosts multiple populations of AGB stars.}

%#######################################################################################

\begin{table}
\caption{Spread in the $c_{\rm y}$ index and average $c_{\rm y}$ uncertainty due to photometric errors.}
\label{Tab:cy-spread}
\centering
\begin{tabular}{lccc}\hline\hline
GC &$c_{\rm y}$ spread  & $c_{\rm y}$ spread  & $c_{\rm y}$ unc \\ 
      &      RGB		&  AGB &  \\
NGC6752 &   0.05 & 0.05 & 0.01\\
NGC1851 &   0.03 &  0.05 & 0.01\\
NGC362  &   0.02  & 0.03 & 0.01\\ 
M92        &   0.02   & 0.02 & 0.01\\
M3          &  0.04    & 0.05 & 0.02\\
\hline
\end{tabular}
\end{table}

%#######################################################################################

%******************************************************************
%                    Discussion
%******************************************************************
\section{Discussion}\label{sect:Disc}
For the first time, we showed the presence of multiple populations along the RGB and AGB based on Str\"omgren photometry. Figure~\ref{Fig:GCs} shows that the $c_{\rm y}$ spread is similar on the AGB and RGB for all clusters. 

The $c_{\rm y}$ spread in NGC\,6752 is the largest in our sample of clusters (see Table\,\ref{Tab:cy-spread}). The large spread in $c_{\rm y}$ is expected as NGC\,6752 shows a large abundance spread in the light elements \citep[e.g. 1\,dex in N, 0.8\,dex in Na and O, 0.5\,dex in Mg and 1.5\,dex in Al][]{Yong2005,Yong2008,Yong2013}. For the other clusters the $c_{\rm y}$ spread varies from 0.018 to 0.037\,mag on the RGB and 0.020 to 0.050 on the AGB. Three out of five clusters show the same $c_{\rm y}$ spreads, within the errors, on the RGB and AGB. The $c_{\rm y}$ spread is different on the AGB compared to the RGB for NGC\,362 and NGC\,1851. This could be due to the smaller sample size compared to the RGB sample. As we have fewer data points on the AGB, the data becomes more susceptible to outliers. The fact that the $c_{\rm y}$ spread is significant compared to the photometric errors and that all clusters have confirmed multiple populations, tells us that the $c_{\rm y}$ spread is due to multiple populations. We conclude that the AGBs in the four clusters under investigation are populated by FG and SG stars. One can thus select FG and SG AGB stars based on their location in the ($c_{\rm y}$,$b$) plane where the FG stars are located on the left side of the RGB and AGB and SG are located on the right side.\\

  Aside for NGC\,6752, we were also able to find three sets of C, N, O, and Na abundances in NGC\,1851, one set for four RGB stars \citep{Yong2015}, a second set for four AGB and five RGB stars \citep{Yong2009} and a third set for two AGB stars from the ESO Phase3 archive system\footnote{\url{http://archive.eso.org/cms.html}}. Comparing these abundances within each sample we identify two populations based on whether or not the N and/or Na abundance are low compared to the sample maximum. We have colour-coded these stars in the bottom right panel of Fig.\,\ref{Fig:GCs} where blue and purple symbols identify stars low in N and Na and red and green symbols stars high in N and Na. Two notes should be made with respect to NGC\,1851: 
  
  1) {\bf While the majority of GCs have homogeneous s-process elements and C$+$N$+$O abundance,} NGC\,1851 {\bf hosts}   two {\bf main} populations: one anomalous population enriched in {\bf C$+$N$+$O} and s-process elements and one so-called normal population characterised by solar s-process elements {\bf \citep{Yong2009, Yong2015, Carretta2011b, Marino2014b}}\footnote{{\bf The presence of star-to-star variations in the overall C$+$N$+$O  abundance in NGC\,1851 has been challenged by \citet{Villanova2010} who have analysed spectra of this cluster and concluded that NGC\,1851 has homogeneous [(C$+$N$+$O)/Fe]. However, the conclusion by \citet{Villanova2010} is in contrast with photometric observations of NGC\,1851 \citep{Milone2008,Milone2017,Han2009,Lardo2012}. Although \citet{Milone2008} showed that the splitting of the SGB can be explained with an age difference of about 1Gyr, the most likely explanation is that the split SGB and RGB in the CMD are the result of a second population with enhanced C$+$N$+$O \citep[e.g.][]{Sbordone2011,Cassisi2008,Ventura2009}. }}.
{\bf Intriguingly, both s-rich and s-normal stars are not consistent with a simple population and host sub-populations with different light element abundances \citep[e.g.][]{Lardo2012,Marino2014b}, where s-rich stars have, on average, higher Na and N content than s-normal stars \citep[e.g.][]{Yong2008,Lardo2012}.}
We here identify two stellar groups based on the N and Na abundances. The identified stars can thus belong to the two main populations in NGC\,1851 and/or to subpopulations making up the normal population. 

2) The middle part of the AGB shows a clear split, possibly as a result of multiple populations. Interestingly we find that the stars to the right of the AGB fiducial (purple triangles) have solar barium abundances.
We thus seem to find that the right side of the AGB is populated by the normal population. Unfortunately, we did not find any information on Ba for stars on the left side of the AGB fiducial and hence we cannot conclude that the left side of the AGB is populated by the anomalous population. It would however, be interesting to see whether the double AGB is due to abundance differences in C+N+O and s-process elements, and/or FG/SG stars with only differences in the abundances of light elements. We therefore provide a catalogue (see Table\,\ref{Tab:Catalogue} in the appendix) of the identified AGB and RGB stars for this and the three other clusters in the hope that they will be useful when selecting AGB and RGB stars for future spectroscopic analyses. \\

We now turn back to NGC\,6752 and compare the observations with stellar evolution theory. From stellar evolution theory we expect that not all stars reach the AGB, in particular horizontal branch stars with {\bf $M\lesssim0.5-0.6M_{\odot}$} do not ascent the AGB \citep{Gratton2010}. According to \citet{Milone2013}, NGC\,6752 is characterised by three chemical populations, a FG that is made up of stars with normal He and low Na ($Y\sim0.25$, [Na/Fe]$\sim-0.03$), and a SG that is made up of stars enhanced in He and Na. This SG can be split up into stars with a normal He and Na enhancement (intermediate population, $\Delta Y = 0.01$ and [Na/Fe]$\sim0.26$) and stars strongly enhanced in He and Na (extreme population, $\Delta Y = 0.03$ and [Na/Fe]$\sim0.61$). From the observed abundances we see that stars of the extreme stellar population with Y~0.28-29 do not ascend the AGB, while stars in the remaining two populations ($Y\sim0.25$ and $Y\sim0.26$) are present along the AGB. This finding is completely in line with \citet{Lapenna2016} but also shows that \citet{Campbell2013} was correct in stating that stars with $Y=0.285$ do not reach the AGB. \\
The observations are in accordance with the synthetic HB simulations by \citet{Cassisi2014}, which considers three stellar populations with $Y=$0.24, 0.25 and 0.27 to reproduce the observed AGB-to-RGB ratio, HB morphology,  HB magnitude,  and  HB colour  distribution  without invoking exceptional mass loss during the HB phase.
 The behaviour can be explained as follows:\\

According to the outline in \citet{Gratton2010}, horizontal branch stars with mass $M\lesssim 0.5-0.6M_{\odot}$ do not ascend the AGB. These so-called AGB-manqu{\'e} stars \citep{Greggio1990} have been found in several clusters \citep[see e.g. the case of NGC\,2808:][]{Castellani2006}. {\bf Similarly,  if, for a given value of the initial He abundance, the mass along the RGB is efficient enough to reduce the mass of the star below the critical value for the He-core flash, the He-flash does not occur. Subsequently, the star leaves the RGB before reaching its tip and populates the blue hook of the HB before populating the white-dwarf cooling sequence\footnote{\bf The referee also pointed out that it is also possible that during the cooling, the H-burning shell present in the tiny residual envelope undergoes a He-flash along the WD sequence. A subsequent mixing episode then ensures that these stars end 
on the HB blue hook too.}. Hence these stars are called RGB-manqu{\'e} and do not ascend the AGB either.}

Some indications of this outline can perhaps be noted when looking at the data for NGC\,6752. We have marked the extreme population on the RGB in Fig.\,\ref{Fig:NGC6752-NGC1851} with a small yellow bullet within the blue diamonds giving the Na-rich stars in NGC\,6752. Interestingly, it seems that the tip of the RGB is not populated with extreme stars. Likewise on the AGB, no intermediate SG stars seem to to reach luminosities as high as FG on the AGB. It is possible that this is just a result of small number statistics. Nevertheless we would urge the spectroscopic community to investigate this further as the spectroscopic information can be used to test stellar evolution predictions, which tell us that the extreme/intermediate population produces AGB/RGB-manqu\'e objects. \\

{\bf  Several works showed that the position of a star along the HB depends on its mass} and that the mass in turn is correlated with the chemical composition through the difference in He abundance amongst the different populations {\bf (e.g.\,D'Antona et al\,2002; Marino et al.\,2011).
  It is then expected that some He-rich HB stars, which are extreme SG stars, do not reach the AGB while SG HB stars with less-extreme chemical compositions can reach the AGB. In some cases, SG stars can terminate their AGB evolution before reaching very bright luminosities (e.g.\,Cassisi et al.\,2014 and references therein). This scenario is consistent with our results on AGB stars in NGC\,6752.
 }

\begin{acknowledgements}
{\bf We are grateful to the anonymous referee for several suggestions that have improved the quality of this manuscript.}
  P.G. acknowledges support from grant no. 2011- 5042 from the Swedish Research Council. S.F acknowledge the grant The New Milky Way from the Knut and Alice Wallenberg Foundation.  P.G. and S.F. acknowledges support from the  Swedish National Space Board. AS acknowledges support from MINECO (ESP2015-66134-R) and Generalitat de Catalunya (SGR2014-1458). {\bf L.C gratefully acknowledge support from the Australian Research Council (grants DP150100250, FT160100402). A.\,P.\,M.\, acknowledges support by the Australian Research Council through Discovery Early Career Researcher Award DE150101816. }
\end{acknowledgements}

\bibliographystyle{aa}
\bibliography{allreferences_Multiple_pop}

\begin{thebibliography}{71}
\expandafter\ifx\csname natexlab\endcsname\relax\def\natexlab#1{#1}\fi

\bibitem[{{Anderson} {et~al.}(2006){Anderson}, {Bedin}, {Piotto}, {Yadav}, \&
  {Bellini}}]{Anderson2006b}
{Anderson}, J., {Bedin}, L.~R., {Piotto}, G., {Yadav}, R.~S., \& {Bellini}, A.
  2006, \aap, 454, 1029

\bibitem[{{Anderson} \& {King}(2006)}]{Anderson2006a}
{Anderson}, J. \& {King}, I.~R. 2006, {PSFs, Photometry, and Astronomy for the
  ACS/WFC}, Tech. rep.

\bibitem[{{{\'A}rnad{\'o}ttir} {et~al.}(2010){{\'A}rnad{\'o}ttir}, {Feltzing},
  \& {Lundstr{\"o}m}}]{Arnadottir2010}
{{\'A}rnad{\'o}ttir}, A.~S., {Feltzing}, S., \& {Lundstr{\"o}m}, I. 2010, \aap,
  521, A40

\bibitem[{{Bastian}(2015)}]{Bastian2015}
{Bastian}, N. 2015, ArXiv e-prints:1510.01330

\bibitem[{{Bastian} {et~al.}(2013){Bastian}, {Cabrera-Ziri}, {Davies}, \&
  {Larsen}}]{Bastian2013}
{Bastian}, N., {Cabrera-Ziri}, I., {Davies}, B., \& {Larsen}, S.~S. 2013,
  \mnras, 436, 2852

\bibitem[{{Brown} {et~al.}(2016){Brown}, {Cassisi}, {D'Antona}, {Salaris},
  {Milone}, {Dalessandro}, {Piotto}, {Renzini}, {Sweigart}, {Bellini},
  {Ortolani}, {Sarajedini}, {Aparicio}, {Bedin}, {Anderson}, {Pietrinferni}, \&
  {Nardiello}}]{Brown2016}
{Brown}, T.~M., {Cassisi}, S., {D'Antona}, F., {et~al.} 2016, \apj, 822, 44

\bibitem[{{Calamida} {et~al.}(2007){Calamida}, {Bono}, {Stetson}, {Freyhammer},
  {Cassisi}, {Grundahl}, {Pietrinferni}, {Hilker}, {Primas}, {Richtler},
  {Romaniello}, {Buonanno}, {Caputo}, {Castellani}, {Corsi}, {Ferraro},
  {Iannicola}, \& {Pulone}}]{Calamida2007}
{Calamida}, A., {Bono}, G., {Stetson}, P.~B., {et~al.} 2007, \apj, 670, 400

\bibitem[{{Campbell} {et~al.}(2013){Campbell}, {D'Orazi}, {Yong},
  {Constantino}, {Lattanzio}, {Stancliffe}, {Angelou}, {Wylie-de Boer}, \&
  {Grundahl}}]{Campbell2013}
{Campbell}, S.~W., {D'Orazi}, V., {Yong}, D., {et~al.} 2013, \nat, 498, 198

\bibitem[{{Campbell} {et~al.}(2006){Campbell}, {Lattanzio}, \&
  {Elliott}}]{Campbell2006}
{Campbell}, S.~W., {Lattanzio}, J.~C., \& {Elliott}, L.~M. 2006, \memsai, 77,
  864

\bibitem[{{Carretta} {et~al.}(2011{\natexlab{a}}){Carretta}, {Bragaglia},
  {Gratton}, {D'Orazi D'Orazi}, \& {Lucatello}}]{Carretta2011}
{Carretta}, E., {Bragaglia}, A., {Gratton}, R., {D'Orazi D'Orazi}, V., \&
  {Lucatello}, S. 2011{\natexlab{a}}, \aap, 535, A121

\bibitem[{{Carretta} {et~al.}(2009{\natexlab{a}}){Carretta}, {Bragaglia},
  {Gratton}, \& {Lucatello}}]{Carretta2009a}
{Carretta}, E., {Bragaglia}, A., {Gratton}, R., \& {Lucatello}, S.
  2009{\natexlab{a}}, \aap, 505, 139

\bibitem[{{Carretta} {et~al.}(2009{\natexlab{b}}){Carretta}, {Bragaglia},
  {Gratton}, {Lucatello}, {Catanzaro}, {Leone}, {Bellazzini}, {Claudi},
  {D'Orazi}, {Momany}, {Ortolani}, {Pancino}, {Piotto}, {Recio-Blanco}, \&
  {Sabbi}}]{Carretta2009b}
{Carretta}, E., {Bragaglia}, A., {Gratton}, R.~G., {et~al.} 2009{\natexlab{b}},
  \aap, 505, 117

\bibitem[{{Carretta} {et~al.}(2007){Carretta}, {Bragaglia}, {Gratton},
  {Lucatello}, \& {Momany}}]{Carretta2007}
{Carretta}, E., {Bragaglia}, A., {Gratton}, R.~G., {Lucatello}, S., \&
  {Momany}, Y. 2007, \aap, 464, 927

\bibitem[{{Carretta} {et~al.}(2011{\natexlab{b}}){Carretta}, {Lucatello},
  {Gratton}, {Bragaglia}, \& {D'Orazi}}]{Carretta2011b}
{Carretta}, E., {Lucatello}, S., {Gratton}, R.~G., {Bragaglia}, A., \&
  {D'Orazi}, V. 2011{\natexlab{b}}, \aap, 533, A69

\bibitem[{{Casagrande} {et~al.}(2014){Casagrande}, {Silva Aguirre}, {Stello},
  {Huber}, {Serenelli}, {Cassisi}, {Dotter}, {Milone}, {Hodgkin}, {Marino},
  {Lund}, {Pietrinferni}, {Asplund}, {Feltzing}, {Flynn}, {Grundahl}, {Nissen},
  {Sch{\"o}nrich}, {Schlesinger}, \& {Wang}}]{Casagrande2014a}
{Casagrande}, L., {Silva Aguirre}, V., {Stello}, D., {et~al.} 2014, \apj, 787,
  110

\bibitem[{{Cassisi} {et~al.}(2008){Cassisi}, {Salaris}, {Pietrinferni},
  {Piotto}, {Milone}, {Bedin}, \& {Anderson}}]{Cassisi2008}
{Cassisi}, S., {Salaris}, M., {Pietrinferni}, A., {et~al.} 2008, \apjl, 672,
  L115

\bibitem[{{Cassisi} {et~al.}(2014){Cassisi}, {Salaris}, {Pietrinferni}, {Vink},
  \& {Monelli}}]{Cassisi2014}
{Cassisi}, S., {Salaris}, M., {Pietrinferni}, A., {Vink}, J.~S., \& {Monelli},
  M. 2014, \aap, 571, A81

\bibitem[{{Castellani} {et~al.}(2006){Castellani}, {Iannicola}, {Bono},
  {Zoccali}, {Cassisi}, \& {Buonanno}}]{Castellani2006}
{Castellani}, V., {Iannicola}, G., {Bono}, G., {et~al.} 2006, \aap, 446, 569

\bibitem[{{D'Antona} {et~al.}(2002){D'Antona}, {Caloi}, {Montalb{\'a}n},
  {Ventura}, \& {Gratton}}]{Dantona2002}
{D'Antona}, F., {Caloi}, V., {Montalb{\'a}n}, J., {Ventura}, P., \& {Gratton},
  R. 2002, \aap, 395, 69

\bibitem[{{D'Antona} {et~al.}(2016){D'Antona}, {Vesperini}, {D'Ercole},
  {Ventura}, {Milone}, {Marino}, \& {Tailo}}]{Dantona2016}
{D'Antona}, F., {Vesperini}, E., {D'Ercole}, A., {et~al.} 2016, \mnras, 458,
  2122

\bibitem[{{Decressin} {et~al.}(2007){Decressin}, {Meynet}, {Charbonnel},
  {Prantzos}, \& {Ekstr{\"o}m}}]{Decressin2007a}
{Decressin}, T., {Meynet}, G., {Charbonnel}, C., {Prantzos}, N., \&
  {Ekstr{\"o}m}, S. 2007, \aap, 464, 1029

\bibitem[{{Denissenkov} \& {Hartwick}(2014)}]{Denissenkov2014}
{Denissenkov}, P.~A. \& {Hartwick}, F.~D.~A. 2014, \mnras, 437, L21

\bibitem[{{D'Orazi} {et~al.}(2015){D'Orazi}, {Gratton}, {Angelou}, {Bragaglia},
  {Carretta}, {Lattanzio}, {Lucatello}, {Momany}, {Sollima}, \&
  {Beccari}}]{Dorazi2015}
{D'Orazi}, V., {Gratton}, R.~G., {Angelou}, G.~C., {et~al.} 2015, \mnras, 449,
  4038

\bibitem[{{Dorman} {et~al.}(1993){Dorman}, {Rood}, \& {O'Connell}}]{Dorman1993}
{Dorman}, B., {Rood}, R.~T., \& {O'Connell}, R.~W. 1993, \apj, 419, 596

\bibitem[{{Dotter} {et~al.}(2015){Dotter}, {Ferguson}, {Conroy}, {Milone},
  {Marino}, \& {Yong}}]{Dotter2015}
{Dotter}, A., {Ferguson}, J.~W., {Conroy}, C., {et~al.} 2015, \mnras, 446, 1641

\bibitem[{{Garc{\'{\i}}a-Hern{\'a}ndez}
  {et~al.}(2015){Garc{\'{\i}}a-Hern{\'a}ndez}, {M{\'e}sz{\'a}ros}, {Monelli},
  {Cassisi}, {Stetson}, {Zamora}, {Shetrone}, \&
  {Lucatello}}]{GarciaHernandez2015}
{Garc{\'{\i}}a-Hern{\'a}ndez}, D.~A., {M{\'e}sz{\'a}ros}, S., {Monelli}, M.,
  {et~al.} 2015, \apjl, 815, L4

\bibitem[{{Gratton} {et~al.}(2012){Gratton}, {Carretta}, \&
  {Bragaglia}}]{Gratton2012}
{Gratton}, R.~G., {Carretta}, E., \& {Bragaglia}, A. 2012, A\&A Review, 20, 50

\bibitem[{{Gratton} {et~al.}(2010){Gratton}, {D'Orazi}, {Bragaglia},
  {Carretta}, \& {Lucatello}}]{Gratton2010}
{Gratton}, R.~G., {D'Orazi}, V., {Bragaglia}, A., {Carretta}, E., \&
  {Lucatello}, S. 2010, \aap, 522, A77

\bibitem[{{Greggio} \& {Renzini}(1990)}]{Greggio1990}
{Greggio}, L. \& {Renzini}, A. 1990, \apj, 364, 35

\bibitem[{{Grundahl} {et~al.}(1999){Grundahl}, {Catelan}, {Landsman},
  {Stetson}, \& {Andersen}}]{Grundahl1999}
{Grundahl}, F., {Catelan}, M., {Landsman}, W.~B., {Stetson}, P.~B., \&
  {Andersen}, M.~I. 1999, \apj, 524, 242

\bibitem[{{Grundahl} {et~al.}(2002){Grundahl}, {Stetson}, \&
  {Andersen}}]{Grundahl2002}
{Grundahl}, F., {Stetson}, P.~B., \& {Andersen}, M.~I. 2002, \aap, 395, 481

\bibitem[{{Grundahl} {et~al.}(1998){Grundahl}, {Vandenberg}, \&
  {Andersen}}]{Grundahl1998}
{Grundahl}, F., {Vandenberg}, D.~A., \& {Andersen}, M.~I. 1998, \apjl, 500,
  L179

\bibitem[{{Han} {et~al.}(2009){Han}, {Lee}, {Joo}, {Sohn}, {Yoon}, {Kim}, \&
  {Lee}}]{Han2009}
{Han}, S.-I., {Lee}, Y.-W., {Joo}, S.-J., {et~al.} 2009, \apjl, 707, L190

\bibitem[{{Ivans} {et~al.}(2001){Ivans}, {Kraft}, {Sneden}, {Smith}, {Rich}, \&
  {Shetrone}}]{Ivans2001}
{Ivans}, I.~I., {Kraft}, R.~P., {Sneden}, C., {et~al.} 2001, \aj, 122, 1438

\bibitem[{{Johnson} {et~al.}(2015){Johnson}, {McDonald}, {Pilachowski},
  {Mateo}, {Bailey}, {Cordero}, {Zijlstra}, {Crane}, {Olszewski}, {Shectman},
  \& {Thompson}}]{Johnson2015}
{Johnson}, C.~I., {McDonald}, I., {Pilachowski}, C.~A., {et~al.} 2015, \aj,
  149, 71

\bibitem[{{Lapenna} {et~al.}(2016){Lapenna}, {Lardo}, {Mucciarelli}, {Salaris},
  {Ferraro}, {Lanzoni}, {Massari}, {Stetson}, {Cassisi}, \&
  {Savino}}]{Lapenna2016}
{Lapenna}, E., {Lardo}, C., {Mucciarelli}, A., {et~al.} 2016, \apjl, 826, L1

\bibitem[{{Lardo} {et~al.}(2012){Lardo}, {Milone}, {Marino}, {Mucciarelli},
  {Pancino}, {Zoccali}, {Rejkuba}, {Carrera}, \& {Gonzalez}}]{Lardo2012}
{Lardo}, C., {Milone}, A.~P., {Marino}, A.~F., {et~al.} 2012, \aap, 541, A141

\bibitem[{{Lardo} {et~al.}(2017){Lardo}, {Salaris}, {Savino}, {Donati},
  {Stetson}, \& {Cassisi}}]{Lardo2017}
{Lardo}, C., {Salaris}, M., {Savino}, A., {et~al.} 2017, \mnras, 466, 3507

\bibitem[{{MacLean} {et~al.}(2016){MacLean}, {Campbell}, {De Silva},
  {Lattanzio}, {D'Orazi}, {Simpson}, \& {Momany}}]{MacLean2016}
{MacLean}, B.~T., {Campbell}, S.~W., {De Silva}, G.~M., {et~al.} 2016, \mnras,
  460, L69

\bibitem[{{Mallia}(1978)}]{Mallia1978}
{Mallia}, E.~A. 1978, \aap, 70, 115

\bibitem[{{Marino} {et~al.}(2015){Marino}, {Milone}, {Karakas}, {Casagrande},
  {Yong}, {Shingles}, {Da Costa}, {Norris}, {Stetson}, {Lind}, {Asplund},
  {Collet}, {Jerjen}, {Sbordone}, {Aparicio}, \& {Cassisi}}]{Marino2015}
{Marino}, A.~F., {Milone}, A.~P., {Karakas}, A.~I., {et~al.} 2015, \mnras, 450,
  815

\bibitem[{{Marino} {et~al.}(2014{\natexlab{a}}){Marino}, {Milone}, {Przybilla},
  {Bergemann}, {Lind}, {Asplund}, {Cassisi}, {Catelan}, {Casagrande},
  {Valcarce}, {Bedin}, {Cort{\'e}s}, {D'Antona}, {Jerjen}, {Piotto},
  {Schlesinger}, {Zoccali}, \& {Angeloni}}]{Marino2014}
{Marino}, A.~F., {Milone}, A.~P., {Przybilla}, N., {et~al.} 2014{\natexlab{a}},
  \mnras, 437, 1609

\bibitem[{{Marino} {et~al.}(2014{\natexlab{b}}){Marino}, {Milone}, {Yong},
  {Dotter}, {Da Costa}, {Asplund}, {Jerjen}, {Mackey}, {Norris}, {Cassisi},
  {Sbordone}, {Stetson}, {Weiss}, {Aparicio}, {Bedin}, {Lind}, {Monelli},
  {Piotto}, {Angeloni}, \& {Buonanno}}]{Marino2014b}
{Marino}, A.~F., {Milone}, A.~P., {Yong}, D., {et~al.} 2014{\natexlab{b}},
  \mnras, 442, 3044

\bibitem[{{Marino} {et~al.}(2011){Marino}, {Villanova}, {Milone}, {Piotto},
  {Lind}, {Geisler}, \& {Stetson}}]{Marino2011}
{Marino}, A.~F., {Villanova}, S., {Milone}, A.~P., {et~al.} 2011, \apjl, 730,
  L16

\bibitem[{{Marino} {et~al.}(2008){Marino}, {Villanova}, {Piotto}, {Milone},
  {Momany}, {Bedin}, \& {Medling}}]{Marino2008}
{Marino}, A.~F., {Villanova}, S., {Piotto}, G., {et~al.} 2008, \aap, 490, 625

\bibitem[{{Milone}(2015)}]{Milone2015c}
{Milone}, A.~P. 2015, \mnras, 446, 1672

\bibitem[{{Milone} {et~al.}(2008){Milone}, {Bedin}, {Piotto}, {Anderson},
  {King}, {Sarajedini}, {Dotter}, {Chaboyer}, {Mar{\'{\i}}n-Franch},
  {Majewski}, {Aparicio}, {Hempel}, {Paust}, {Reid}, {Rosenberg}, \&
  {Siegel}}]{Milone2008}
{Milone}, A.~P., {Bedin}, L.~R., {Piotto}, G., {et~al.} 2008, \apj, 673, 241

\bibitem[{{Milone} {et~al.}(2014){Milone}, {Marino}, {Bedin}, {Piotto},
  {Cassisi}, {Dieball}, {Anderson}, {Jerjen}, {Asplund}, {Bellini}, {Brogaard},
  {Dotter}, {Giersz}, {Heggie}, {Knigge}, {Rich}, {van den Berg}, \&
  {Buonanno}}]{Milone2014b}
{Milone}, A.~P., {Marino}, A.~F., {Bedin}, L.~R., {et~al.} 2014, \mnras, 439,
  1588

\bibitem[{{Milone} {et~al.}(2013){Milone}, {Marino}, {Piotto}, {Bedin},
  {Anderson}, {Aparicio}, {Bellini}, {Cassisi}, {D'Antona}, {Grundahl},
  {Monelli}, \& {Yong}}]{Milone2013}
{Milone}, A.~P., {Marino}, A.~F., {Piotto}, G., {et~al.} 2013, \apj, 767, 120

\bibitem[{{Milone} {et~al.}(2015{\natexlab{a}}){Milone}, {Marino}, {Piotto},
  {Bedin}, {Anderson}, {Renzini}, {King}, {Bellini}, {Brown}, {Cassisi},
  {D'Antona}, {Jerjen}, {Nardiello}, {Salaris}, {Marel}, {Vesperini}, {Yong},
  {Aparicio}, {Sarajedini}, \& {Zoccali}}]{Milone2015b}
{Milone}, A.~P., {Marino}, A.~F., {Piotto}, G., {et~al.} 2015{\natexlab{a}},
  \mnras, 447, 927

\bibitem[{{Milone} {et~al.}(2015{\natexlab{b}}){Milone}, {Marino}, {Piotto},
  {Renzini}, {Bedin}, {Anderson}, {Cassisi}, {D'Antona}, {Bellini}, {Jerjen},
  {Pietrinferni}, \& {Ventura}}]{Milone2015a}
{Milone}, A.~P., {Marino}, A.~F., {Piotto}, G., {et~al.} 2015{\natexlab{b}},
  \apj, 808, 51

\bibitem[{{Milone} {et~al.}(2017){Milone}, {Piotto}, {Renzini}, {Marino},
  {Bedin}, {Vesperini}, {D'Antona}, {Nardiello}, {Anderson}, {King}, {Yong},
  {Bellini}, {Aparicio}, {Barbuy}, {Brown}, {Cassisi}, {Ortolani}, {Salaris},
  {Sarajedini}, \& {van der Marel}}]{Milone2017}
{Milone}, A.~P., {Piotto}, G., {Renzini}, A., {et~al.} 2017, \mnras, 464, 3636

\bibitem[{{Monelli} {et~al.}(2013){Monelli}, {Milone}, {Stetson}, {Marino},
  {Cassisi}, {del Pino Molina}, {Salaris}, {Aparicio}, {Asplund}, {Grundahl},
  {Piotto}, {Weiss}, {Carrera}, {Cebri{\'a}n}, {Murabito}, {Pietrinferni}, \&
  {Sbordone}}]{Monelli2013}
{Monelli}, M., {Milone}, A.~P., {Stetson}, P.~B., {et~al.} 2013, \mnras, 431,
  2126

\bibitem[{{Nardiello} {et~al.}(2015){Nardiello}, {Piotto}, {Milone}, {Marino},
  {Bedin}, {Anderson}, {Aparicio}, {Bellini}, {Cassisi}, {D'Antona}, {Hidalgo},
  {Ortolani}, {Pietrinferni}, {Renzini}, {Salaris}, {Marel}, \&
  {Vesperini}}]{Nardiello2015}
{Nardiello}, D., {Piotto}, G., {Milone}, A.~P., {et~al.} 2015, \mnras, 451, 312

\bibitem[{{Norris} {et~al.}(1981){Norris}, {Cottrell}, {Freeman}, \& {Da
  Costa}}]{Norris1981}
{Norris}, J., {Cottrell}, P.~L., {Freeman}, K.~C., \& {Da Costa}, G.~S. 1981,
  \apj, 244, 205

\bibitem[{{Pancino} {et~al.}(2010){Pancino}, {Rejkuba}, {Zoccali}, \&
  {Carrera}}]{Pancino2010}
{Pancino}, E., {Rejkuba}, M., {Zoccali}, M., \& {Carrera}, R. 2010, \aap, 524,
  A44

\bibitem[{{Piotto} {et~al.}(2015){Piotto}, {Milone}, {Bedin}, {Anderson},
  {King}, {Marino}, {Nardiello}, {Aparicio}, {Barbuy}, {Bellini}, {Brown},
  {Cassisi}, {Cool}, {Cunial}, {Dalessandro}, {D'Antona}, {Ferraro}, {Hidalgo},
  {Lanzoni}, {Monelli}, {Ortolani}, {Renzini}, {Salaris}, {Sarajedini}, {van
  der Marel}, {Vesperini}, \& {Zoccali}}]{Piotto2015}
{Piotto}, G., {Milone}, A.~P., {Bedin}, L.~R., {et~al.} 2015, \aj, 149, 91

\bibitem[{{Renzini} {et~al.}(2015){Renzini}, {D'Antona}, {Cassisi}, {King},
  {Milone}, {Ventura}, {Anderson}, {Bedin}, {Bellini}, {Brown}, {Piotto}, {van
  der Marel}, {Barbuy}, {Dalessandro}, {Hidalgo}, {Marino}, {Ortolani},
  {Salaris}, \& {Sarajedini}}]{Renzini2015}
{Renzini}, A., {D'Antona}, F., {Cassisi}, S., {et~al.} 2015, \mnras, 454, 4197

\bibitem[{{Sbordone} {et~al.}(2011){Sbordone}, {Salaris}, {Weiss}, \&
  {Cassisi}}]{Sbordone2011}
{Sbordone}, L., {Salaris}, M., {Weiss}, A., \& {Cassisi}, S. 2011, \aap, 534,
  A9

\bibitem[{{Smith} \& {Norris}(1993)}]{Smith1993}
{Smith}, G.~H. \& {Norris}, J.~E. 1993, \aj, 105, 173

\bibitem[{{Str{\"o}mgren}(1963)}]{Stromgren1963}
{Str{\"o}mgren}, B. 1963, \qjras, 4, 8

\bibitem[{{Sweigart} \& {Gross}(1976)}]{Sweigart1976}
{Sweigart}, A.~V. \& {Gross}, P.~G. 1976, \apjs, 32, 367

\bibitem[{{Ventura} {et~al.}(2009){Ventura}, {Caloi}, {D'Antona}, {Ferguson},
  {Milone}, \& {Piotto}}]{Ventura2009}
{Ventura}, P., {Caloi}, V., {D'Antona}, F., {et~al.} 2009, \mnras, 399, 934

\bibitem[{{Villanova} {et~al.}(2010){Villanova}, {Geisler}, \&
  {Piotto}}]{Villanova2010}
{Villanova}, S., {Geisler}, D., \& {Piotto}, G. 2010, \apjl, 722, L18

\bibitem[{{Villanova} {et~al.}(2012){Villanova}, {Geisler}, {Piotto}, \&
  {Gratton}}]{Villanova2012}
{Villanova}, S., {Geisler}, D., {Piotto}, G., \& {Gratton}, R.~G. 2012, \apj,
  748, 62

\bibitem[{{Wang} {et~al.}(2016){Wang}, {Primas}, {Charbonnel}, {Van der
  Swaelmen}, {Bono}, {Chantereau}, \& {Zhao}}]{Wang2016}
{Wang}, Y., {Primas}, F., {Charbonnel}, C., {et~al.} 2016, \aap, 592, A66

\bibitem[{{Yong} {et~al.}(2009){Yong}, {Grundahl}, {D'Antona}, {Karakas},
  {Lattanzio}, \& {Norris}}]{Yong2009}
{Yong}, D., {Grundahl}, F., {D'Antona}, F., {et~al.} 2009, \apjl, 695, L62

\bibitem[{{Yong} {et~al.}(2008){Yong}, {Grundahl}, {Johnson}, \&
  {Asplund}}]{Yong2008}
{Yong}, D., {Grundahl}, F., {Johnson}, J.~A., \& {Asplund}, M. 2008, \apj, 684,
  1159

\bibitem[{{Yong} {et~al.}(2005){Yong}, {Grundahl}, {Nissen}, {Jensen}, \&
  {Lambert}}]{Yong2005}
{Yong}, D., {Grundahl}, F., {Nissen}, P.~E., {Jensen}, H.~R., \& {Lambert},
  D.~L. 2005, \aap, 438, 875

\bibitem[{{Yong} {et~al.}(2015){Yong}, {Grundahl}, \& {Norris}}]{Yong2015}
{Yong}, D., {Grundahl}, F., \& {Norris}, J.~E. 2015, \mnras, 446, 3319

\bibitem[{{Yong} {et~al.}(2013){Yong}, {Mel{\'e}ndez}, {Grundahl}, {Roederer},
  {Norris}, {Milone}, {Marino}, {Coelho}, {McArthur}, {Lind}, {Collet}, \&
  {Asplund}}]{Yong2013}
{Yong}, D., {Mel{\'e}ndez}, J., {Grundahl}, F., {et~al.} 2013, \mnras, 434,
  3542

\end{thebibliography}

\begin{appendix} 
\section{Selection rules}
\begin{landscape}
\begin{table}
\caption{AGB and RGB star selection rules.}
\label{Tab:selection}
\centering
\begin{tabular}{l|c|c|c|c|c}\hline\hline
GC & HB $y$  & $m_1$ & CMD & RGB & AGB   \\
 \hline
M3 & 15.5 & $-0.1< m_1 < 0.2$ or & $(v-y)$ vs $y$ & $\Delta(v-y) > 0.06$ & $\Delta(v-y)$ \tablefootmark{a} $ < 0.06$ \\
      &      & $1.075(b-y)-0.5 < m_1 < 1.075(b-y)-0.3$ & & & \\ \hline
M92 & 15 &  $m_1 > 0$ & $(v-y)$ vs $y$ &  $\Delta(v-y)$ \tablefootmark{b} $ < 0.03$ & $1 < (v-y) < 1.45$, $c_{\rm y} > -0.1$ \\
	&	&  		     & and $c_{\rm y}$ vs $b$      &  $\Delta y$ \tablefootmark{b} $ < 0.2$       & $\Delta(v-y)$ \tablefootmark{a} $ < 0.03$ and $\Delta y$ \tablefootmark{a} $ < 0.15$ \\ \hline
NGC\,362 & 15 & $0.175(u-y)-0.3 < m_1 < 0.175(u-y)-0.19$ & $(v-y)$ vs $y$ & $\Delta(v-y) < 0.04$ and $\Delta y < 0.18$ and & $-2(v-y)+16.9< y<-2(v-y)+17.2$ \\
		&	&								    &			      & $-2(v-y)+17.2 < y$ and $(v-y) < 2.5$ & $1.2 < (v-y) < 1.8$ \\ \hline
NGC\,1851 & 15.9 & $0.175(u-y)-0.3< m_1 < 0.175(u-y)-0.175$ & $(v-y)$ vs $y$ & $\Delta(v-y) < 0.04$ and $\Delta y < 0.18$ & $7.06c_{\rm y}+17.0<b$ \\
		&	    &									& and $c_{\rm y}$ vs $b$	& and $(v-y) < 1.7$ & and $-3(v-y)+19.6<y$ \\ 
\hline
\end{tabular}
\tablefoot{
\tablefoottext{a}{Distance to a fiducial going through the by-eye identified AGB in the CMD.}
\tablefoottext{b}{Distance to a fiducial going through the by-eye identified RGB in the CMD.}
}

\end{table}
\end{landscape}

\section{Catalogue}

\begin{table}
\caption{RGB and AGB catalogue based on the Grundahl photometry. Full table available on CDS.}
\label{Tab:Catalogue}
\centering
\begin{tabular}{lccccc}\hline\hline
GC & id & Ra & Dec & $y$ mag  & Type \\ 
 \hline
NGC\,362 & 813 &  15.834377 & -70.899370 & 14.92 & RGB \\
       		&  866 & 15.827797 & -70.898116 & 13.93 & RGB \\
                 & 1086 & 15.871138 & -70.892643 & 14.79 & RGB \\
                 & 1305 & 15.787733 & -70.889132 & 14.77 & RGB \\
                 & 2195 & 15.758909 & -70.875510 & 14.39 & RGB \\

\hline
\end{tabular}
\end{table}

\begin{table}
\caption{RGB and AGB catalogue for M3. Full table available on CDS.}
\label{Tab:M3}
\centering
\begin{tabular}{ccccccc}\hline\hline
id & Ra & Dec & $u$ & $v$ & $b$ & $y$ \\ 
 \hline
RGB-1 &    205.7058305  &   28.3981320 & 17.244 & 16.015 & 15.182 & 14.589 \\
RGB-2 &    205.6732088  &   28.3189091 & 17.203 & 15.265 & 13.655 & 12.660 \\
RGB-3 &    205.6639402  &   28.4663573 & 16.837 & 15.110 & 13.873 & 13.048 \\
RGB-4 &    205.6480061  &   28.3328729 & 16.766 & 15.346 & 14.311 & 13.609 \\
RGB-5 &    205.6432670  &   28.4169339 & 17.182 & 16.012 & 15.239 & 14.635 \\
RGB-6 &    205.6338469  &   28.3839849 & 17.067 & 15.906 & 15.165 & 14.589 \\
\hline
\end{tabular}
\end{table}

\end{appendix}
\end{document}